\documentclass[11pt,a4paper]{article}
\pdfoutput=1
\usepackage{jheppub}
\usepackage{graphicx,placeins}
\usepackage{wrapfig}

\usepackage[utf8]{inputenc}

\def\lsim{\raise0.3ex\hbox{$<$\kern-0.75em\raise-1.1ex\hbox{$\sim$}}}
\def\gsim{\raise0.3ex\hbox{$>$\kern-0.75em\raise-1.1ex\hbox{$\sim$}}}

\newcommand{\bee}{\begin{equation}}
\newcommand{\ee}{\end{equation}}

\begin{document}

\preprint{
{
\flushright
ITP-Budapest 657\\  
CPT-P004-2012\\     
WUB/12-02\\       
}
}

\title{High-precision scale setting in lattice QCD}

\collaboration{Budapest-Marseille-Wuppertal collaboration}

\author[a]{Szabolcs Bors\'anyi,}
\author[a,b]{Stephan D\"urr,}
\author[a,b,c]{Zolt\'an Fodor,}
\author[a]{Christian Hoelbling,}
\author[c]{S\'andor D. Katz,}
\author[a,b]{Stefan Krieg,}
\author[a]{Thorsten Kurth,}
\author[d]{Laurent Lellouch,}
\author[a,b]{Thomas Lippert,}
\author[a]{Craig McNeile}
\author[a]{and K\'alm\'an K. Szab\'o}

%
%
\emailAdd{borsanyi@uni-wuppertal.de,durr@itp.unibe.ch,fodor@bodri.elte.hu,
hch@physik.uni-wuppertal.de,katz@bodri.elte.hu,s.krieg@fz-juelich.de,
thorsten.kurth@uni-wuppertal.de,lellouch@cpt.univ-mrs.fr,
th.lippert@fz-juelich.de,mcneile@uni-wuppertal.de,szaboka@general.elte.hu}

\affiliation[a]{Bergische Universit\"at Wuppertal, Gaussstr.\,20, D-42119 Wuppertal, Germany.}
\affiliation[b]{J\"ulich Supercomputing Centre, Forschungszentrum J\"ulich, D-52425 J\"ulich, Germany.}
\affiliation[c]{Institute for Theoretical Physics, E\"otv\"os University, H-1117 Budapest, Hungary.}
\affiliation[d]{Centre de Physique Th\'eorique, \footnote{\scriptsize CPT is research unit UMR 7332 of the CNRS, of Aix-Marseille U. and of U. Sud Toulon-Var; it is affiliated with the CNRS' research federation FRUMAM (FR 2291).} CNRS, Aix-Marseille U. and U. Sud Toulon-Var, F-13288 Marseille, France.}

\date{\today}

\abstract{ Scale setting is of central importance in lattice QCD. It is 
required to predict dimensional quantities in physical units. Moreover, 
it determines the relative lattice spacings of computations performed 
at different values of the bare coupling, and this is needed for 
extrapolating results into the continuum. 
Thus, we calculate a new quantity, $w_0$, for setting the scale in lattice QCD,
which is based on the Wilson flow like the scale $t_0$ (M.~Luscher, JHEP 1008
(2010) 071).
It is cheap and straightforward to implement and compute. In particular, it
does not involve the delicate fitting of correlation functions at asymptotic
times. It typically can be determined on the few per-mil level. We compute its
continuum extrapolated value in $2+1$-flavor QCD for physical and non-physical
pion and kaon masses, to allow for mass-independent scale setting even away
from the physical mass point.  We demonstrate its robustness by computing it
with two very different actions (one of them with staggered, the other with
Wilson fermions) and by showing that the results agree for physical quark
masses in the continuum limit.
}


\maketitle

\section{Introduction}

\begin{figure*}[b]
\centerline{
\includegraphics[scale=0.6]{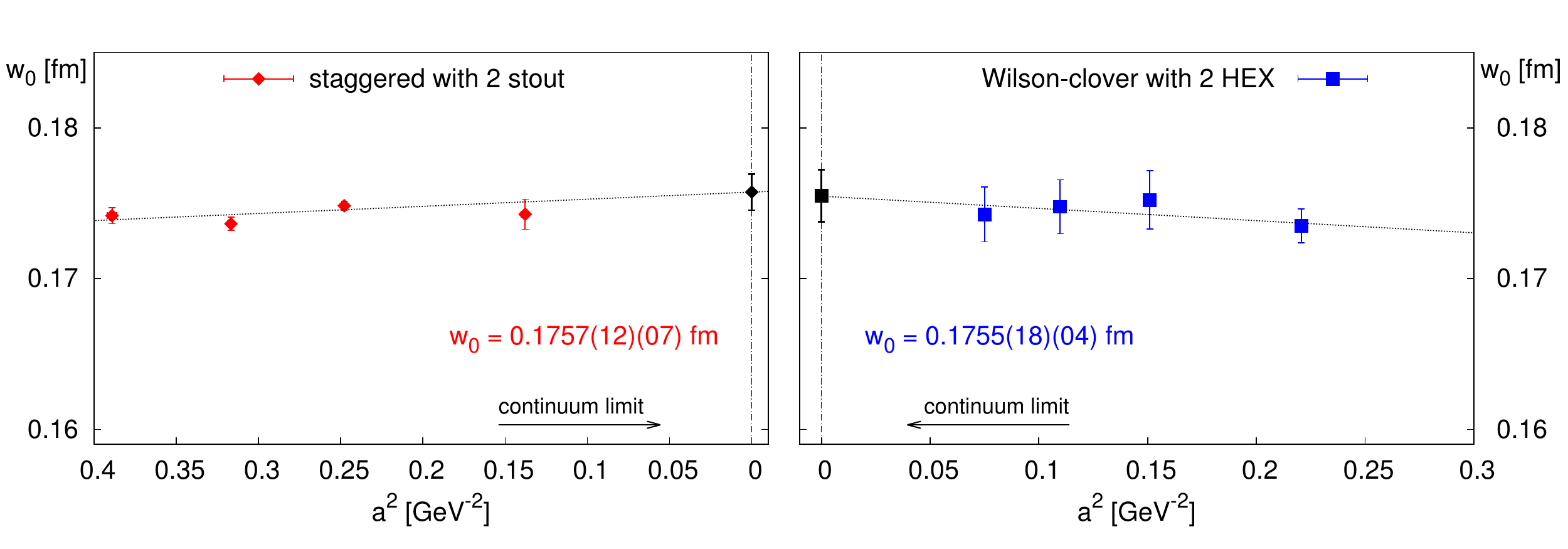}
}
\caption{\label{final_plain} 
Representative continuum extrapolations of the $w_0$ scale, at the physical 
mass point. The values at different lattice spacings are obtaind by 
using the Wilson flow described below. The continuum limit values on 
the plots are results from our final, full analyses. The results 
obtained with the two very different actions {(staggered fermions on the 
left and Wilson fermions on the right panel)} are in good agreement and 
the overall uncertainties are very small. 
}
\end{figure*}

Quantum chromodynamics (QCD) is a theory with few parameters: the quark 
masses and an overall scale. To determine the latter, one has to 
compute a dimensionful quantity or an observable at a known energy, and 
adjust the overall scale of the theory to reproduce the corresponding 
experimental measurement. In lattice calculations this is equivalent to 
fixing the lattice spacing $a$. The lattice spacing is determined by 
calculating a dimensionful quantity $Q$ --for definiteness, $Q$ is 
chosen to be of mass dimension one here-- such as the mass of the Omega 
baryon, $M_\Omega$, or the pion decay constant, $F_\pi$, and by 
relating its lattice value to its experimental value through $a= (a 
Q)^{latt} / Q^{expt}$. In principle any dimensionful quantity can be 
used. However, it is clear that the quality of any dimensionful prediction
from the lattice can only be 
as good as the quality of the determination of the overall scale. This 
statement is particularly relevant now that lattice QCD results with 
errors below 2\% are beginning to be reported.

Besides the overall scale in physical units, it is also important to 
accurately determine the relative lattice spacings of simulations 
performed at different values of the bare coupling in order to carry 
out a continuum extrapolation. Independent calculations can be compared 
too, if dimensionful quantities are expressed in units of a well 
measured quantity. For this purpose it may be useful to consider an 
observable which is not directly measured in experiment, but which is 
particularly simple to compute with high accuracy. Moreover, if this 
observable is computed accurately in physical units once, its value can 
be used in all subsequent lattice calculations to fix their overall 
scales.

One popular observable of this type is the Sommer scale, $r_0$, 
introduced nearly two decades ago~\cite{Sommer:1993ce}. More recently, 
MILC has found it more convenient to consider the related scale 
$r_1$~\cite{Bernard:2000gd}. One of the advantages of these scales is 
that they are based on the calculation of the static potential from 
gauge fields: they do not require the costly computation of quark 
propagators, as do observables such as $M_\Omega$ or $F_\pi$. However, 
the determination of the potential requires a delicate study of the 
asymptotic time behavior of Wilson loops and the calculation of $r_0$ 
or $r_1$ from the potential is a much more complicated analysis 
(see~\cite{Donnellan:2010mx} for a recent example) than, say, fitting 
the mass of a particle, such as the pion, from a measured correlator. 
The introduction of HYP smearing~\cite{Hasenfratz:2001hp} on the gauge 
links has reduced the problem of the poor signal-to-noise ratio of 
Wilson loops, but the calculation of $r_0$ or $r_1$ remains 
non-trivial. 
{These subtleties might be at least partially responsible 
for the tension within present determinations of 
the Sommer scale between \cite{Aoki:2010dy} and
\cite{Davies:2009tsa,Bazavov:2011aa}, and for a similar 
tension between recent and present determinations 
\cite{Follana:2007uv,Bazavov:2011aa}. While these differences are only 
on the two standard deviation level they have an important impact on 
the search for new physics in the leptonic decays of the $D_s$ meson, 
as discussed in~\cite{Kronfeld:2009cf}}.

In this paper we propose an alternative to the $r_0$ scale: the $w_0$ 
scale.
The method is based on the Wilson flow. The Wilson flow was considered 
in the context of trivializing maps by Luscher \cite{Luscher:2009eq}. 
It was studied earlier by Narayanan and Neuberger 
\cite{Narayanan:2006rf} in a different context, too. Its important 
renormalization properties were clarified in 
\cite{Luscher:2010iy,Luscher:2011bx}. Its application to scale setting,
which we build upon, was suggested recently in \cite{Luscher:2010iy}.

 The $w_0$ scale keeps the advantages of the Sommer scale (i.e. 
no expensive fermion inversion is needed). However, it is easier to 
determine with high precision, since it requires neither a fitting of 
the asymptotic time behavior of correlation functions nor fighting 
signal-to-noise issues.

Before discussing the method in detail (see Section \ref{method}), we 
present our results for the $w_0$ scale in physical units. These values 
can henceforth be used to determine the lattice spacing in physical 
units in $N_f=2+1$ lattice QCD calculations. (The dependence of $w_0$ 
on $N_f$ should be studied before it is used to set the scale in 
simulations with $N_f\ne 2+1$.) We performed two independent 
calculations of $w_0$, both based on simulations with pion masses all 
the way down to its physical value {and below.} The first uses our 2HEX 
smeared Wilson fermion ensembles 
\cite{Durr:2010aw,Durr:2010vn,Durr:2011ap} and the second, 2-stout 
smeared staggered simulations 
\cite{Aoki:2006we,Aoki:2006br,Aoki:2009sc,Borsanyi:2010bp,Borsanyi:2010cj,Endrodi:2011gv,Borsanyi:2011sw}. 
In both cases $w_0$ is interpolated to the physical quark masses as 
well as extrapolated into the continuum. The $\Omega$ mass is used to 
convert these scales to physical units ({with our smeared actions 
hadron mass ratios show very small cutoff effects 
\cite{Durr:2008rw,Durr:2008zz})}. Representative continuum limits (see 
below) are displayed in Fig.~\ref{final_plain}, where the staggered and 
Wilson results are shown on the l.h.s. and r.h.s., respectively. The 
plot indicates that $w_0$ has cutoff effects similar to $M_\Omega$, 
resulting in a very mild continuum scaling, and that the uncertainties 
on the extrapolated value are very small. Moreover, the staggered and 
Wilson results are in good agreement and the precisions reached with 
the two actions are on the same level.  We quote the Wilson result, 
which does not rely on the ``rooting'' of the fermion determinant, as 
our final result:
\bee
w_0=0.1755(18)(04)~\mathrm{fm}\label{result_wilson}\ ,
\ee
where the first error is statistical and the second is systematic. Note 
that the overall uncertainty is 1\%, most of which is statistical. 
Furthemore, the statistical error in the dimensionless quantity 
$w_0M_\Omega$ comes dominantly from $aM_\Omega$. Thus, the error on 
$w_0/a$ itself is subdominant, typically on the level of a few per mil 
or less. This fact makes $w_0/a$ a particularly attractive candidate to 
set the relative scale between simulations for continuum extrapolations 
and for comparing calculations from different groups. Another 
interesting application is the determination of the ratio of the 
lattice spacings for anisotropic actions \cite{Borsanyi:2011xx}. As a 
side product we also compute a related quantity $(t_0)^{1/2}$ suggested 
in \cite{Luscher:2010iy} (though on the same set of configurations its 
{\em relative} systematic error is {four times} larger than that of 
$w_0$).

{This paper is organized as follows. After this introductory section we 
discuss the scale setting method based on the Wilson (and Symanzik) 
flow in Section \ref{method}. The next two sections 
(\ref{wilson},\ref{staggered}) deal with our results obtained with 
Wilson and with staggered fermions, respectively. Section \ref{finiteV} 
discusses two possible problems, namely finite volume effects and 
autocorrelations for the flow. Section \ref{conclusions} presents the 
final results and concludes. In order to make the practical application 
of the scale setting procedure presented here easier, the method is 
implemented in the CHROMA software system \cite{Edwards:2004sx}. In addition, 
along with this paper we submit two codes to the arXiv, both written in 
C, as ancillary files. The first ({\tt wilson\_flow.c}) determines the 
Wilson (and Symanzik) flow. It was written emphasizing readability over 
speed. It works both for isotropic and anisotropic 
\cite{Borsanyi:2011xx} lattice actions. The second one ({\tt 
w0\_scale.c}) uses the output of {\tt wilson\_flow.c} or that of CHROMA 
to determine the scale $w_0/a$ and its statistical uncertainty.
}

\section{The scales $(t_0)^{1/2}$ and $w_0$\label{method}}

The scale setting method can be summarized as follows. Following
the strategy of Ref.~\cite{Luscher:2010iy} we calculate the Wilson flow,
that is we integrate infinitesimal gauge-field
smearing steps up to a scale $t$, whose units are inverse 
mass-squared. The smearing is performed until a well-chosen 
dimensionless observable reaches a specified value. The universal 
``flow time,'' $t=t_0$, at which this happens can then be used to set 
the scale on the original lattices.

Integrating the infinitesimal smearing steps is equivalent to finding 
the solution to the flow equation 
\cite{Narayanan:2006rf,Luscher:2010iy}:
\begin{equation}
\dot{V}_t=Z(V_t)V_t,\quad V_0=U 
\label{wflow_equation} 
\end{equation} 
where $V_t$ are the gauge links at flow time $t$ and $U$ are the original 
gauge links.  {In \cite{Luscher:2010iy,Luscher:2011bx}, where the 
Wilson action is used, $Z(V_t)$ is the derivative of the plaquette action
and the corresponding flow is called the Wilson flow.  As it can be seen
from Eq.~(\ref{wflow_equation}) an infinitesimal change of the link
variable is obtained by the product of the link variable itself and the
sum of the staples around
it. Thus, for the present case the flow is generated by infinitesimal
stout-smearing steps. As a consequence the action decreases and the gauge
field is getting smoother.  For improved gauge actions, one can take
$Z(V_t)$ to be the algebra-valued 
derivative of the gauge action.}

To obtain the scale $t_0$, it is suggested in \cite{Luscher:2010iy} to 
integrate the flow and to compute $t^2 \langle E(t)\rangle$ as a 
function of $t$, $t_0$ being the flow time where $t^2 
\langle E(t)\rangle$ reaches $0.3$.  Here $\langle E(t)\rangle$ is the 
expectation value of the continuum-like action density 
$G^a_{\mu\nu}(t)G^a_{\mu\nu}(t)/4$, where $G^a_{\mu\nu}(t)$ is a 
lattice version of the chromoelectric field-strength tensor at flow 
time $t$. Here we use the usual clover-leaf definition for this tensor. 
Note that $t^2 \langle E\rangle$ {turned out to be approximately 
proportional} to $t$ for large flow times, a similar observation was made
for the pure gauge theory in Ref.~\cite{Luscher:2010iy}.

\begin{figure}
\centerline{\includegraphics[scale=0.65]{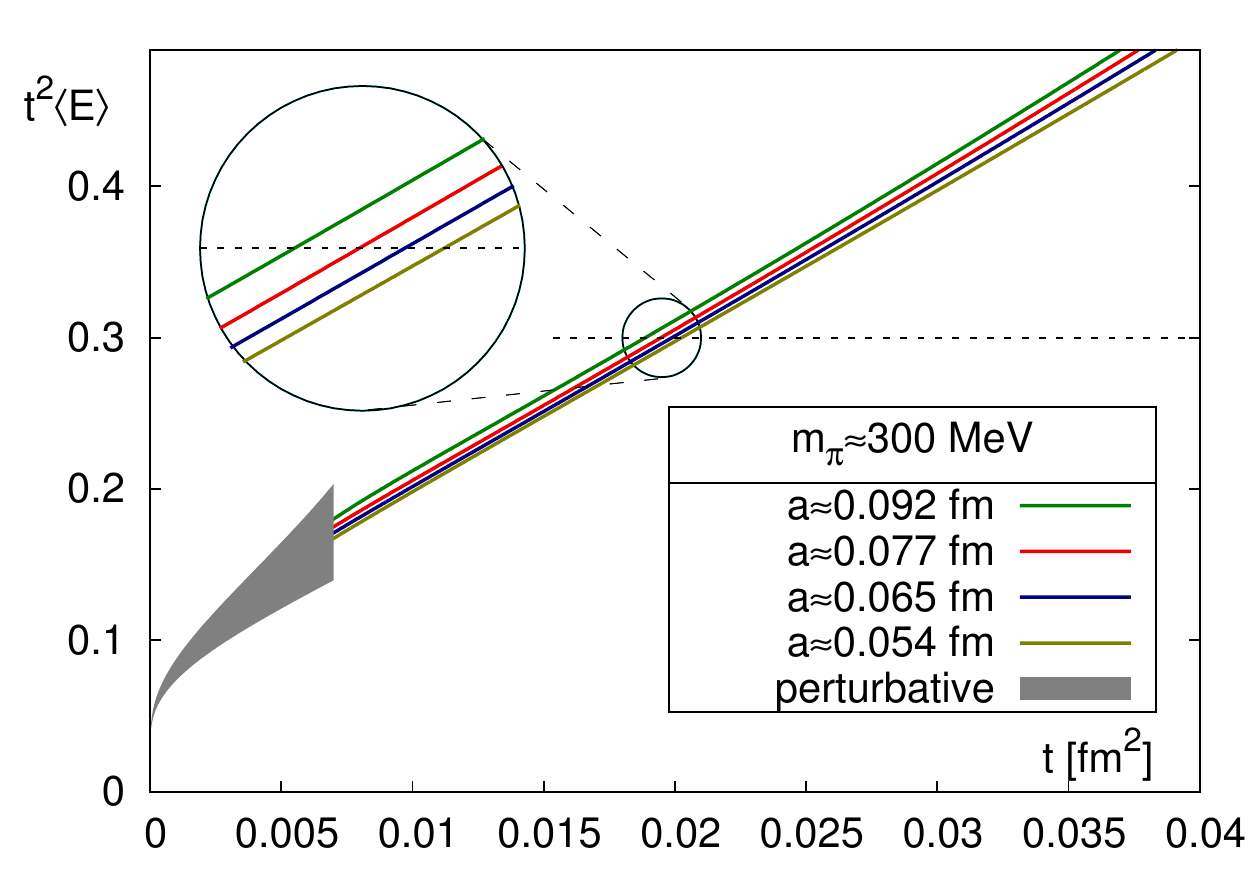}}
\caption{\label{fig_compare} {Analogously to the Wilson flow in Fig. 2 
of \cite{Luscher:2010iy} here we show the {Symanzik} flow, which can be 
used to define the $w_0$ scale and the $t_0^{1/2}$ scale proposed by 
\cite{Luscher:2010iy}. These flows are obtained using our $N_f$=2+1 
Wilson fermion simulations. For our four lattice spacings we have 
runs close to $M_\pi$$\approx$300~MeV (the individual pion masses are 
somewhat different). The perturbative expectation is shown by the gray 
band. Its width indicates the uncertainty in $\Lambda_{QCD}$ using two 
representative values from the literature (c.f. Refs. 
\cite{Shintani:2010ph,McNeile:2010ji}).}} 
\end{figure}

Here we propose to use another, related observable, namely
{
\begin{equation}\label{our_flow} 
W(t)\equiv t\frac{d}{dt}\left\{t^2 \langle E(t)\rangle\right\}
\end{equation} 
and define the $w_0$ scale, via the condition 
\bee
\left. W(t) \right|_{t=w_0^2}=0.3\ .
\ee
}

{The most important reasons for this choice can be summarized as 
follows.} While $t^2 \langle E(t)\rangle$ incorporates information 
about the gauge configurations from all scales larger than ${\cal 
O}(1/\sqrt{t})$ {(thus including scales also around the cutoff)}, 
$W(t)$ mostly depends on scales around ${\cal O}(1/\sqrt{t})$. This is 
an advantage, because the behavior of the flow at small $t\sim a^2$ is 
subject to discretization effects. {Let us illustrate these cutoff 
effects by one example}. The flow $t^2 \langle E(t)\rangle$ starts 
vertically at the origin in the continuum, while it must start 
horizontally for any lattice spacing and for any lattice action. {The 
value of $t^2 \langle E(t)\rangle$ at $t$ is influenced by this cutoff 
effect appearing at small $t$, whereas the derivative $W(t)$ is less 
affected.}  In the original approach, the 
flow times $t$ corresponding to different values of $t^2 \langle 
E(t)\rangle$ yield different relative scales. {Contrary to that,} 
$W(t)$ yields very similar scales when different values are considered 
on the r.h.s.\ of (\ref{our_flow}).  Furthermore, the perturbative 
calculation of \cite{Luscher:2010iy} provides strong evidence that 
$t_0$, and also $w_0$, does not require renormalization.

\begin{figure*}
\centerline{\includegraphics[scale=0.6]{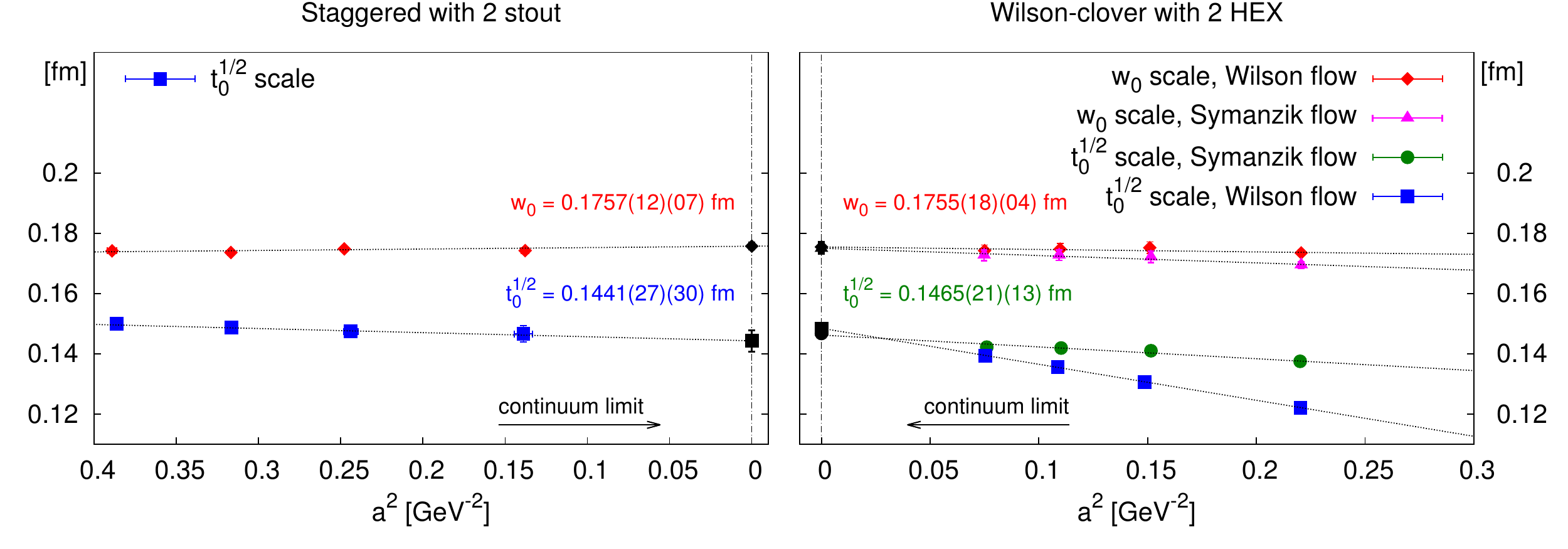}}
\caption{\label{final_plus}
Representative continuum extrapolations of two different scale 
observables ($t_0^{1/2}$ and $w_0$) at physical quark masses, using our 
$N_f=2+1$ staggered (left panel) and Wilson (right panel) simulation 
data. {An illustration of the different cutoff effects is also shown. We 
determined $t_0^{1/2}$ and $w_0$ with Wilson fermions using two 
different flow equations (Wilson and Symanzik flow). The results based 
on the different flow equations show different discretization effects.} 
For the two very different actions the final results are in good 
agreement and the overall uncertainties on the continuum values are 
very small.}
\end{figure*}

\begin{figure*}[b]
\centerline{
\includegraphics[scale=0.6]{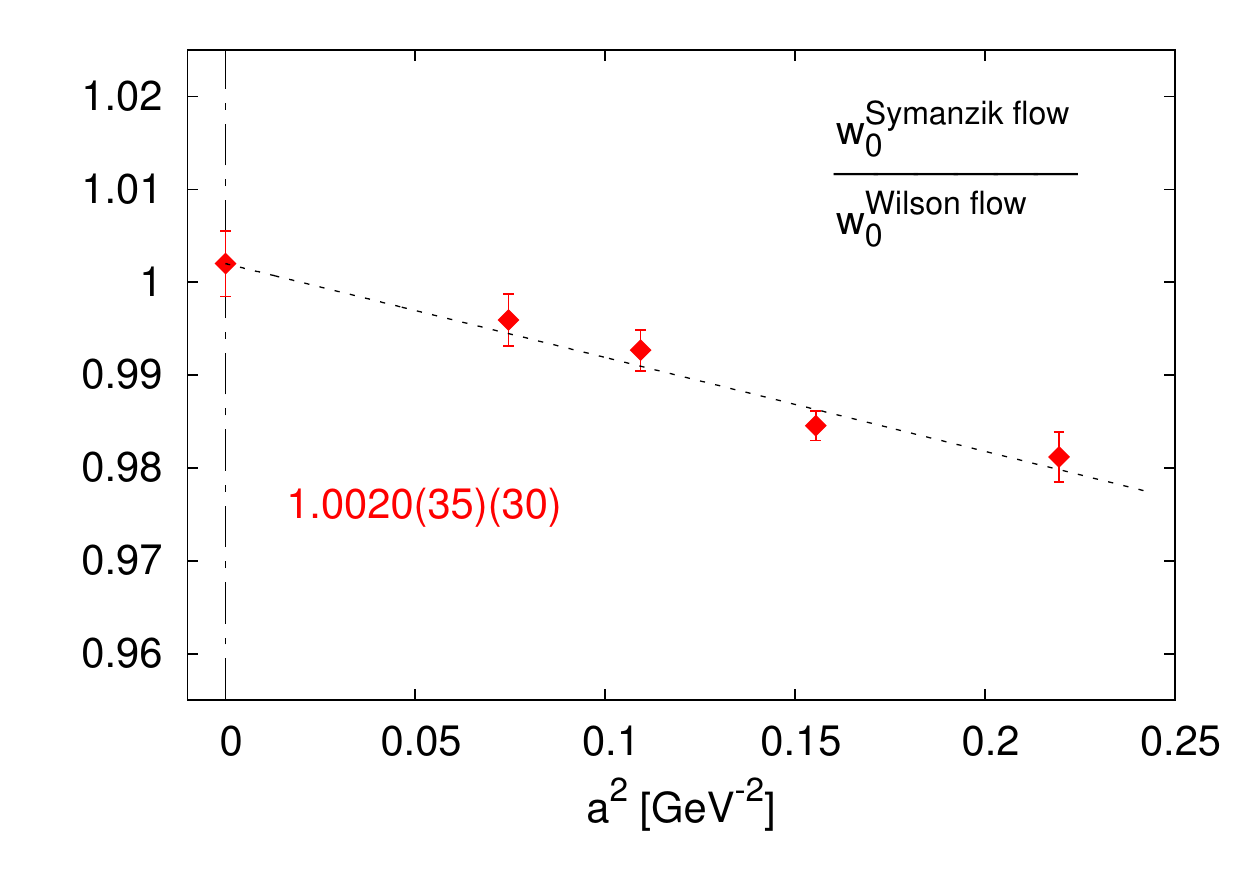}
}
\caption{\label{flow_ratio.pdf}
{Ratio of the $w_0$ scales obtained by Symanzik and Wilson flows
for physical pion and kaon masses. 
}
}
\end{figure*}

\begin{figure*}
\hbox to\hsize{
\includegraphics[scale=0.30]{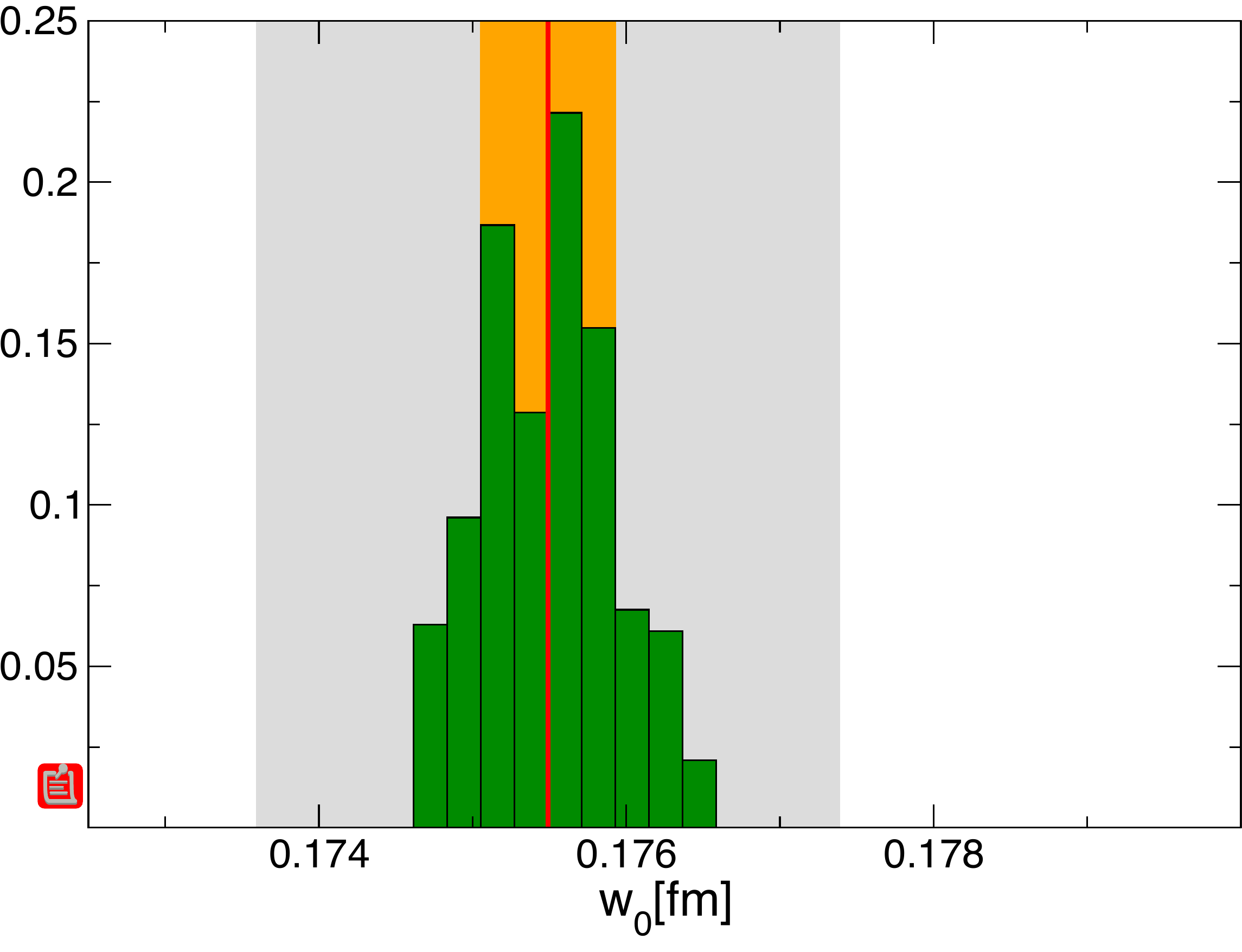}
}
\caption{\label{histogram}Final histogram for the different
analyses used to compute $w_0$ with our 
Wilson fermion simulations. Each entry is weighted by its 
corresponding fit quality. The orange (thinner) band denotes the 
systematic and the gray (thicker) one the {combined (systematic and}
statistical) uncertainties. 
The vertical line depicts the central value. Due to the small width
of the distribution (note the scale), our final result 
(\ref{result_wilson}) is very precise.}
\end{figure*}

Of course one is free to modify the lattice observable used or the flow 
equation (\ref{wflow_equation}) by terms which vanish in the continuum 
limit. For instance, as Section~3 of \cite{Luscher:2010iy} discusses, 
$\langle E(t)\rangle$ can be obtained directly from the sum of the 
plaquettes or from the more symmetric clover definiton of 
$G^a_{\mu\nu}(t)$. Both definitions are acceptable and lead to results 
which must converge to the same continuum limit. One can look at the 
pure $SU(3)$ gauge theory with no quarks and study the flow in units of 
$r_0$. The symmetric definition turns out to be 
advantageous \cite{Luscher:2010iy}, since it results in negligible cutoff
effects, whereas the non-symmetric definition leads to approximately 5\% cutoff
effects at a lattice spacing of 0.1~fm \cite{Luscher:2010iy}. The 
choice of $Z(V_t)$ in (\ref{wflow_equation}) is also only fixed up to 
discretization corrections. The natural choice is to consider the 
algebra-valued derivative of the gauge action used in the simulation 
(cf. Section~6 of \cite{Luscher:2009eq}). In our case this is a 
tree-level Symanzik improved action~\cite{Luscher:1984xn} and we call 
the corresponding flow, the Symanzik flow {(c.f. 
Fig.~\ref{fig_compare})}.  However, the use of the Wilson flow is also 
correct and should give the same continuum limit. This is illustrated 
in Fig.~\ref{final_plus}, with 2HEX Wilson fermions. As can be seen, 
$t_0^{1/2}$ obtained by the Wilson flow has larger cutoff effects than 
the one coming from the Symanzik flow. For $w_0$, both choices are good 
and cutoff effects are around or less than two percent, in the range of 
lattice spacings considered. This is confirmed by looking at the ratio 
of the $w_0$ scales obtained from the Symanzik and the Wilson flows as 
a function of lattice spacing. Fig.~\ref{flow_ratio.pdf} shows that the 
ratio appoaches 1 in the continuum limit, as it should. Here the pion 
and kaon masses are tuned to their physical values (this tuning leads 
to systematic uncertainties included on the points and discussed 
below). For our staggered action both $t_0^{1/2}$ and $w_0$ can be 
obtained with small cutoff effects using the Wilson flow.
Note that in our case it is consistent to use one action for generating
the fields and another one in the gauge flow. The two actions differ only
in higher order in the lattice spacing, thus the derivative of the action
with respect to the link variable will have the same order cutoff effect.
Since in our setup the fermionic sector has already $a^2$ cutoff effects
--at least--, using the Wilson flow instead of the Symanzik flow does not
deteriorate the continuum extrapolation. Obviously, one should use the
same definition for different lattice spacings, if a continuum
extrapolation is carried out.

In fact the $w_0$ scale determined with the Wilson 
flow has tiny cutoff effects for both our Wilson and 
staggered actions. Since integrating the Wilson flow is several times 
faster than working with the Symanzik flow, the former provides a quick 
and straightforward determination of the lattice spacing through the 
$w_0$ scale.

{On a practical level, the flow equation (\ref{wflow_equation}) can be 
efficiently integrated by using the explicit fourth-order Runge-Kutta 
scheme proposed in \cite{Luscher:2010iy}. The links at flow time 
$t+\epsilon$ are obtained from those at flow time $t$ via
\begin{align}\label{runge-kutta}
  X_0&=V_t,\nonumber\\ X_1&=\exp\left(\frac{1}{4}Z_0\right)X_0,\nonumber\\ X_2&=\exp\left(\frac{8}{9}Z_1-\frac{17}{36}Z_0\right)X_1,\nonumber\\ 
V_{t+\epsilon}&=\exp\left(\frac{3}{4}Z_2-\frac{8}{9}Z_1+\frac{17}{36}Z_0\right)X_2,
\end{align}
where $Z_i\equiv\epsilon Z(X_i)$. It turns out that the step size 
$\epsilon$ can be chosen rather coarse, since the total integration error
associated with the 
finite step size scales like $\epsilon^3$ \cite{Luscher:2010iy}. Indeed 
we find that a value of $\epsilon=0.01$ yields finite-step-size errors 
far below the per-mil level, which is negligible for our purposes. 
These findings are in agreement with those of \cite{Luscher:2010iy}.}

\section{$N_f=2+1$ Wilson fermion computation\label{wilson}}

We compute the $w_0$ scale {(and also the $t_0^{1/2}$ scale)} using 
our 2HEX smeared \cite{Capitani:2006ni} Wilson fermion ensembles 
\cite{Durr:2010vn,Durr:2010aw,Durr:2011ap}, dropping the coarsest 
lattice with $\beta=3.31$, {as it appears to be less suited for 
studying various options for flows and/or scale setting procedures.} 
Note that we are still left with four lattice spacings, {which provide 
a safe continuum extrapolation,} and pion masses down to, or even 
below, the physical value.

{ As discussed above, the Symanzik or the Wilson flows are equally 
valid for determining scale observables. Our continuum results for 
these observables agree within systematic errors. In order to reduce 
the uncertainties coming from the continuum extrapolation, one should 
favor the flow which has small cutoff effects. As we illustrated in 
Fig.~\ref{final_plus} the continuum extrapolation of the $w_0$ scale 
with the Symanzik and the Wilson flows are almost equally good (the 
Wilson flow is slightly better). For the $t_0^{1/2}$ scale the Symanzik 
flow gives smaller cutoff effects than the Wilson flow, resulting in  
a factor of two smaller 
systematic error. Thus, for our results (c.f. Fig.~\ref{final_plus}) we 
used the flows with the smaller cutoff effects. Even with this 
favourable choice the {\em relative} systematic error of the $t_0^{1/2}$
scale is still about four times larger than that 
of the $w_0$ scale. This huge difference in accuracy justifies our 
preference for the $w_0$ scale (for the $w_0$ scale Symanzik or Wilson 
flows are similarly good). Moreover, since integrating the Wilson flow 
is several times faster than integrating the Symanzik flow, the best 
way to set the scale is to use $w_0$ determined from the Wilson 
flow.}

As discussed above, we use the $\Omega$ baryon mass to express $w_0$ in 
physical units. Thus, the scale is extracted from $aM_\Omega$ at the 
point where the ratios $(M_\pi/M_\Omega,M_K/M_\Omega)$ acquire their 
physical values, as described in \cite{Durr:2010aw}. We then compute 
$w_0(M_\pi,M_K,a)$ in physical units for each ensemble and perform a 
combined quark-mass interpolation and continuum extrapolation to obtain 
the physical value of $w_0$.

Four different fit functions are used to interpolate to the physical 
mass point in the $M_\pi$--$M_K$ plane. They have the form 
$w_0$=$a_0$+$a_1$$M_\pi^2$+$a_2$$M_K^2$+d(a)+$hoc.$, where $hoc.$ 
stands for higher order contributions in $M_\pi^2$ and/or $M_K^2$. 
Because our fermion action is tree-level improved, the discretization 
corrections, $d(a)$, are chosen to be either proportional to $\alpha_s 
a$ or $a^2$ {(since the continuum extrapolation of $w_0/M_\Omega$ is 
practically constant, essentially no difference is observed between 
these two choices).}

\begin{wraptable}{r}[0mm]{8cm}
\begin{tabular}{lr}\hline
\hline
\small source & relative error $[\%]$\\\hline
\small physical point interpolation & $15$\\
\small $M_\pi$-cut & $40$\\
\small continuum limit & $55$\\
\small spectrum & $55$\\
\small scale & $45$\\
\hline\hline
\end{tabular}
\caption{\label{error_budget}Contributions to the systematic 
uncertainty on $w_0$, as fractions of the total systematic error in \% 
{(rounded to the closest 5\%)}. {The various uncertainties are 
explained in the main text and they are listed in the same order here.}  
Note that the fractions must be added in quadrature and do not sum up 
exactly to 100\% due to correlations and rounding.
}
\end{wraptable}

The various strategies that we apply for the mass extractions, to 
interpolate to the physical point and to extrapolate to the continuum 
limit are all combined to estimate the systematic uncertainties. To 
that end we use 64 different analyses, i.e. the 4 different fit forms 
in the $M_\pi$--$M_K$ plane, 2 pion mass cuts for these fits 
($M_\pi<300\,\mathrm{MeV},\,350\,\mathrm{MeV}$), 2 different scaling 
assumptions in the lattice spacings, 2 fit ranges for extracting $M_K$, 
$M_\pi$ and $M_\Omega$ as well as 2 methods for setting the scale (corresponding 
to different pion mass cuts in the $M_\Omega$-fits, i.e. 
$M_\pi<380\,\mathrm{MeV},\,480\,\mathrm{MeV}$). Each of these analyses 
can be fully justified and can be considered ``the'' final analysis. 
Thus, according to our standard procedure 
\cite{Durr:2008zz,Durr:2010aw,Durr:2010hr,Durr:2011ap}, we construct a 
histogram out of the values obtained for $w_0$, where each one is 
weighted by the corresponding fit quality. We compute the median and 
the central $68\%$ confidence interval of the resulting distribution 
and take these values to be our central value and systematic error, 
respectively (cf. Fig.~\ref{histogram}). A detailed error budget is 
given in Tab.~\ref{error_budget}. The same set of analyses has been 
repeated for the observable based on the Symanzik flow. We found an 
agreement within error (see Fig.~\ref{final_plus}).

The statistical error is computed by repeating the analysis on 2000 
bootstrap samples. Note that the statistical error is much larger than 
the systematic. The statistical error of $w_0/a$ is much smaller than 
that of $aM_\Omega$. Therefore, the error of our $w_0$ in physical 
units is dominated by the statistical uncertainties in $aM_\Omega$. In 
that way, we obtain the result with its systematic and statistical 
errors given in Eq.~(\ref{result_wilson}).

\section{$N_f=2+1$ staggered fermion computation\label{staggered}}

\begin{table}
\begin{center}
\begin{tabular}{lll}
\hline
\hline
$\beta$, scale& $am_s $& $m_s/m_u$ \\
\hline
3.7500 & 0.050254 & 28,14,10\\
$a^{-1}=1.605(6)(3)$ GeV & 0.048 & 27.9,20,10\\
 & 0.040 & 10\\
\hline
3.7920 & 0.05 & 20\\
$a^{-1}=1.778(7)(1)$ GeV & 0.045 & 28,20,14,10\\
 & 0.040 & 20,10\\
\hline
3.8500 & 0.0395 & 27.3,20,14,10\\
$a^{-1}=2.024(18)(7)$ GeV & 0.0388 & 20,14,10\\
 & 0.037 & 10\\
\hline
3.9900 & 0.0283 & 28.15,10,6\\
$a^{-1}=2.684(58)(7)$ GeV & 0.0277 & 14,10,6\\
\hline
\hline
\end{tabular}
\end{center}
\caption{\label{stout-ensemble}
Staggered ensembles used in this analysis.
}
\end{table}

{An interesting test is a comparison of continuum results obtained with 
Wilson and staggered fermions. Therefore, we perform a fully 
independent determination of $w_0$. We consider the 2-step 
stout-smeared staggered fermion action \cite{Morningstar:2003gk} used 
in our thermodynamics studies} 
\cite{Aoki:2006we,Aoki:2006br,Aoki:2009sc,Borsanyi:2010bp,Borsanyi:2010cj,Borsanyi:2011zza}. 
The parameters of the ensembles used here are summarized in 
Tab.~\ref{stout-ensemble}. Note that the pion masses either straddle 
the physical value {(obtained from $M_\pi/M_\Omega$) or even touch it 
within errors. We express all quantities as functions of the bare 
masses for fixed gauge coupling. The $w_0$ scale in lattice units, 
$w_0/a(am_{ud},am_s)$, is then computed for each simulation point as 
described before. These results are then interpolated to the physical 
point. This interpolation is done for every lattice spacing separately. 
Again, we use four functional forms as in the Wilson case. {Since 
there is no additive mass renormalization for staggered fermions and 
the bare quark masses are known exactly,} in the interpolating fits we 
use these masses. Thus, instead of $M_\pi^2$ we use 2$m_{ud}$ and 
instead of $M_K^2$ we use $m_{ud}+m_s$ in the fit functions.} For both the
kaon and pion mass, we use three polynomial fit formulae 
{to describe their quark mass dependence}. Finally, we 
perform a linear continuum extrapolation in $a^2$. 
In order to estimate the cutoff effects we use the four or the three finest
lattice spacings.
We end up with $72$ 
different continuum values for $w_0$, where each one can be weighted by 
the combined goodness of fit. 
Note that we have a much larger statistics for our staggered action than
for the Wilson one.

On a subset of the ensembles used here we have already carried
out a scaling study for $r_0$ in Ref.~\cite{Aoki:2009sc}.
We found that for the same action and lattice spacing range one observes
about 10\% cutoff effect for $r_0$ (see the right panel of Fig. 4
in Ref.~\cite{Aoki:2009sc}, where the scale was set by $f_K$).
We also gave a scaling plot for $M_\Omega$ in the same paper.

\section{Finite volume effects and autocorrelations\label{finiteV}}

\begin{figure}
\centerline{\includegraphics[width=8cm]{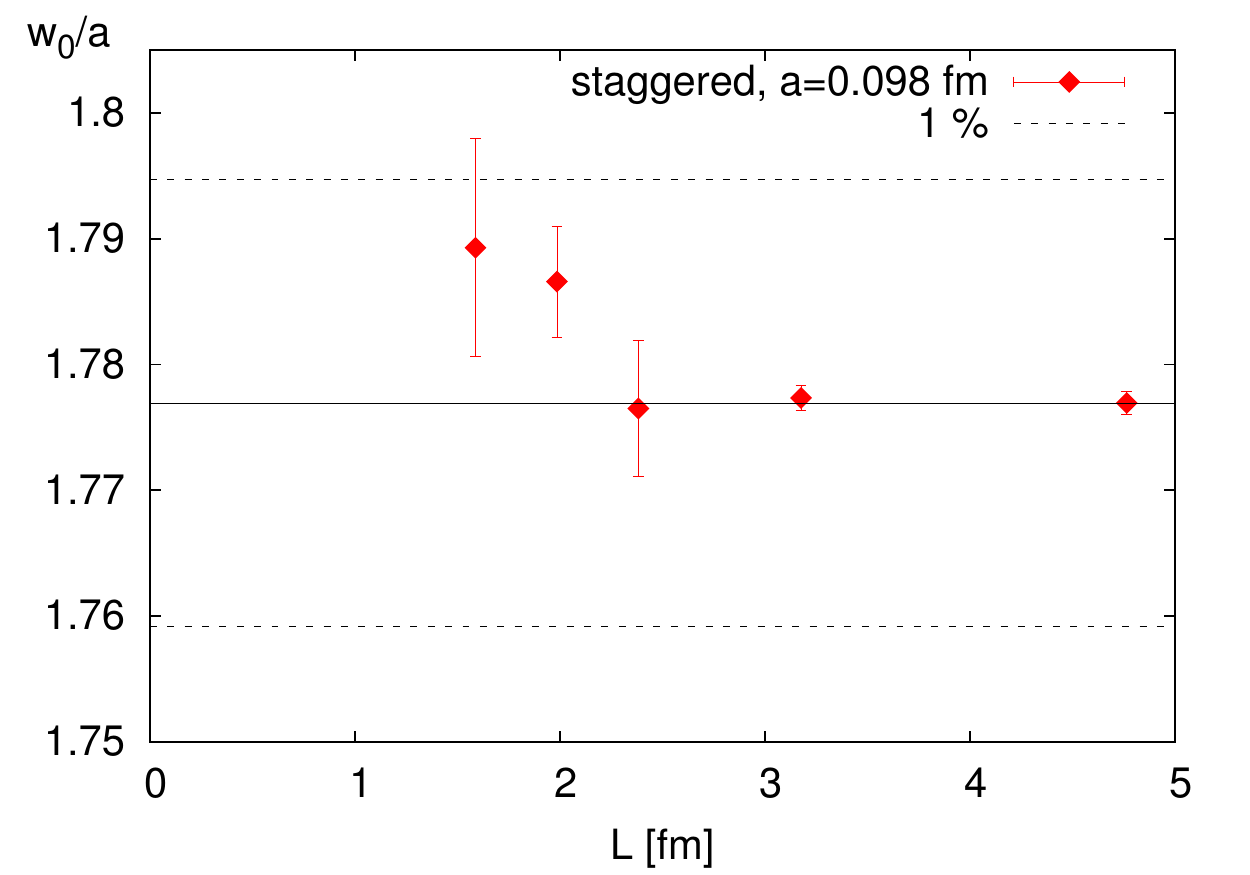}}
\caption{\label{fv_effects} Finite-volume effects in the $w_0$ 
  scale. Here we plot the measured values of $w_0/a$ as a function of
  $L$, obtained with fixed bare quark masses corresponding to the
  physical point.  Notice that it is perfectly feasible to determine
  $w_0/a$ to per-mil precision on the 164 configurations we have on
  our $48^3\times 64$ lattice. Even for boxes of size slightly below
  2~fm, deviations from the large-volume value never exceed 1\%.  }
\end{figure}

At fixed physical volume, finite-volume effects on the Wilson or 
Symanzik flow increase as the flow time increases. It is, therefore, 
important to check that these effects remain small for our choice of 
$w_0$ scale. Fig.~\ref{fv_effects} displays the volume dependence of 
$w_0/a$ on the second finest staggered lattice at physical quark 
masses. {(For this test we used a couple of thousand trajectories for 
each volume.)} It shows that finite-volume effects only become relevant 
for box sizes $\lesssim 2\,\mathrm{fm}$. For lattices larger than that, 
finite volume effects are essentially absent. {Note that for a lattice 
of 2~fm one has $M_\pi L$$=$1.35, which is far smaller than the spatial 
size suggested by the rule of thumb $M_\pi L$$\gsim$4. These tiny 
finite volume effects and the smallness of the error on $w_0/a$ make 
the $w_0$ scale a very attractive intermediate quantity to determine 
the lattice spacing.}

{Since the Wilson/Symanzik flow incorporates more and more information 
from the whole lattice as $t$ increases, it is important to study 
autocorrelations.} Thus, we compute the integrated autocorrelation time 
of the flow as a function of $t$, using the standard methodology 
described in \cite{Wolff:2003sm}.  We do so by looking at several long 
(5000 trajectories) parallel HMC streams on our finest staggered 
lattices or simply analyzing a long HMC stream on one of our finest 
Wilson ensembles \cite{Durr:2010aw}. We find that for lattice spacings 
down to about $a=0.54$~fm, the integrated autocorrelation time of 
$E(t=w_0^2)$ is around or below 50 unit-length trajectories. These 
autocorrelations are taken into account by appropriate binning.

\section{Results and conclusions\label{conclusions}}

We presented a new quantity for setting the scale in lattice QCD 
calculations. Precise determinations of this new $w_0$ scale were 
obtained using Wilson and staggered fermion simulations with lattice 
speacings down to 0.054~fm and average up and down quark masses all the 
way down to, and even below, its physical value. Therefore, we {showed} 
that the $w_0$ scale can be used to reliably determine the lattice 
spacing in physical units in upcoming lattice calculations. Moreover, 
the good agreement between the Wilson and staggered determinations 
{illustrates} the robustness of this scale-setting method {(c.f.
 Fig.~\ref{final_plain}).}

\begin{figure}[h!]
\centerline{\includegraphics[width=12cm]{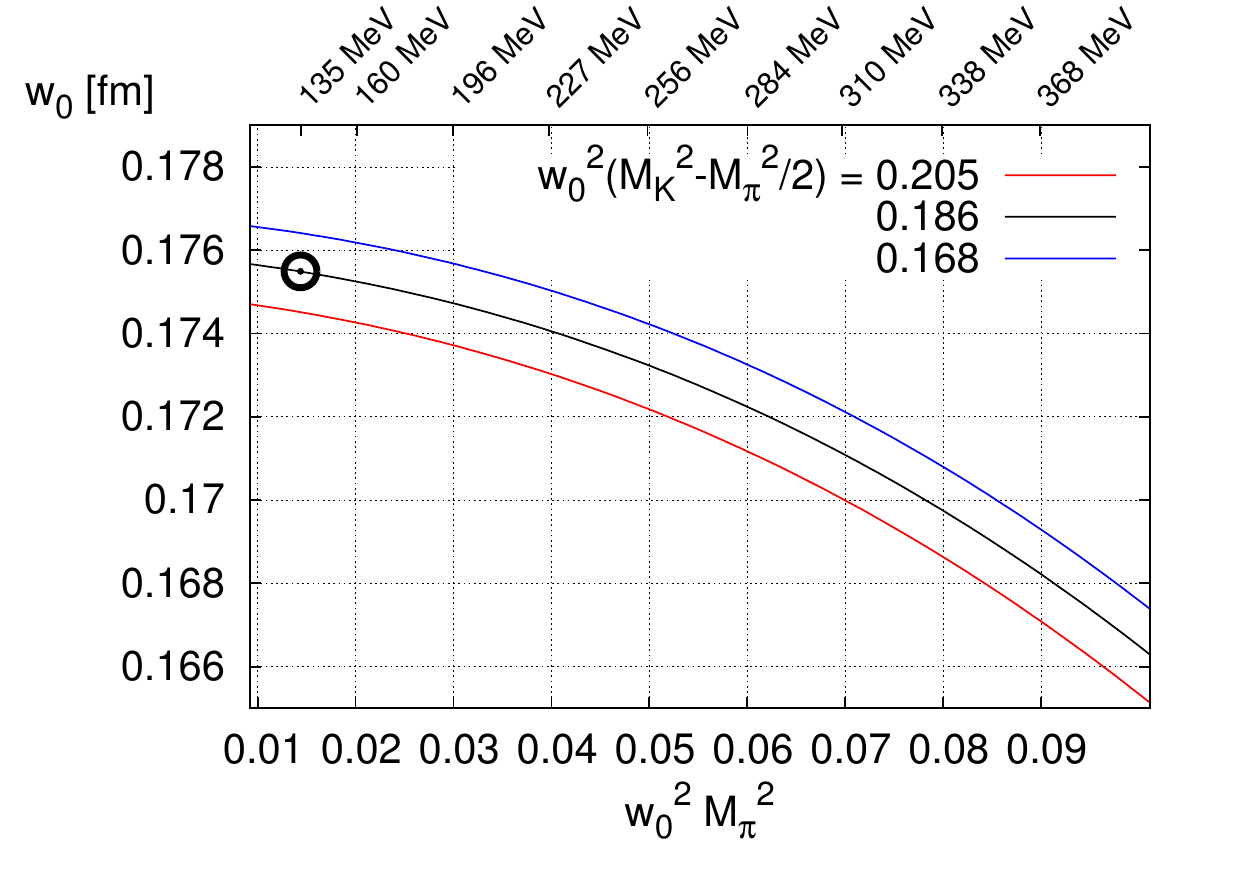}}
\caption{\label{mpidep} The value of the $w_0$ scale at various pion 
and kaon masses in the continuum limit. 
The results are based on our Wilson simulations. The black dot 
represents the physical point. To use the figures one should compute 
$aM_\pi$, $aM_K$ and $w_0/a$. Then one reads off the value of $w_0$ in 
physical units corresponding to the combinations $w_0^2M_\pi^2$ and 
$w_0^2(M_K^2-M_\pi^2/2)$. The 
black curve in the middle corresponds to the physical $M_K^2-M_\pi^2/2$ 
(used to fix $m_s$), whereas the lines below/above correspond to a 10\% 
smaller/larger value of this quantity. As it can be seen the change in 
$w_0$ is small. Changing the pion mass from its physical value to a 
three times larger value (thus, almost an order of magnitude larger 
quark mass) results in about 5\% change in $w_0$. Changing the strange 
quark mass by 10\% means changing $w_0$ on the 0.5\% level. The mass 
independent scale setting prescription is used here.
}
\end{figure}

In Eq.~(\ref{our_flow}) we define the $w_0$ scale as the square root of 
the ``time'' at which the logarithmic derivative of $t^2 \langle 
E(t)\rangle$ reaches 0.3. A larger value, say 0.5, increases the cost 
of integrating the flow, as well as the size of statistical and 
finite-volume errors. A smaller value would probe short-distance 
physics which is more strongly affected by discretization errors. For 
values below 0.1, this becomes a serious concern for coarser lattices. 
The value 0.3 is chosen to be safely away from these two extremes and 
is optimal for modern day simulations, performed with lattice spacings 
in the range $0.05\,\mathrm{fm}\lsim a\lsim 0.1\,\mathrm{fm}$ and 
lattice sizes larger than 2~fm.

The continuum extrapolated value of $w_0$ for the physical point is 
given in Eq.~(\ref{result_wilson}):
\bee
w_0=0.1755(18)(04)~\mathrm{fm}.\nonumber
\ee

For non-physical pion and/or kaon masses the pion and kaon mass 
dependence of $w_0$ is displayed in Fig.~\ref{mpidep}. This figure 
allows to determine the lattice spacing and its uncertainty as follows. 
One measures $M_\pi a$, $M_K a$ and $w_0/a$. Using these three 
quantities one then determines the dimensionless combinations $x=w_0^2 
M_\pi^2$ and $y=w_0^2(M_K^2-M_\pi^2/2)$. These two quantities define a 
point in Fig.~\ref{mpidep}. Reading off the value of $w_0$ in physical 
units and combining it with the computed value of $w_0/a$ gives the 
lattice spacing in fm. {For pion and kaon masses covered by this 
figure the uncertainty of the $w_0$ scale is essentially mass 
independent and its value is 1\% (fully dominated by the statistical error 
of $M_\Omega$).} The same result for $w_0$ and the lattice 
spacing can be obtained by using the following formula:
\begin{equation} w_0= 0.18515 -0.5885x^2 
-0.0497y -0.11xy-1.476x^3 {\pm 18\cdot 10^{-3}(stat)\pm
4\cdot 10^{-3}(sys)}  
\end{equation}
which is valid for $0.01 \lsim x \lsim 0.1$ and $0.165 \lsim y \lsim 
0.205$.

In this paper we have shown that the $w_0$ scale has several advantages 
over other scale setting procedures (most of which are also shared by $t_0$,
though the latter is more sensitive to cut-off effects).
The most important are:
\begin{itemize}
\item $w_0$ is cheap and easy to implement and compute (note that our
reference implementations are publicly available); in particular:

\begin{itemize}
\item $w_0$ does not require the computation of quark propagators;

\item $w_0$ does not involve the delicate fitting of correlation
  functions at asymptotic times.

\end{itemize}

\item the determination of $w_0$ is not only cheap but it 
can be done precisely and reliably; typically one obtains results 
with an accuracy on the few per mil level;

\item the value of $w_0$ in physical units is known for physical and
non-physical quark masses (for mass independent scale setting);

\item our results suggest that $w_0$ depends weakly on quark masses;
in particular, unlike scale setting with $M_\Omega$, even a
  $10\%$ deviation in $m_s$ from its physical value only translates
  into a $\lsim 0.5\%$ change in the $w_0$  scale; 

\item {in the present investigation the most precise and 
fastest method to determine the scale was $w_0$ based on the Wilson 
flow: independently of the type of the flow $w_0$ has quite small 
cutoff effects and consequently small systematic uncertainties, whereas 
integrating the Wilson flow is the fastest among all flow choices.}

\end{itemize}

\textbf{Acknowledgements:} Computations were performed using HPC 
resources from FZ J\"ulich and from GENCI-[IDRIS/CCRT] (grant 52275) 
and on the special purpose QPACE computer and GPU clusters at Wuppertal 
\cite{Egri:2006zm}. This work is supported in part by EU grants I3HP, 
FP7/2007-2013/ERC No. 208740, MRTN-CT-2006-035482 (FLAVIAnet), DFG 
grants FO 502/2, SFB-TR 55, by CNRS grants GDR 2921 and PICS 4707.

\bibliographystyle{JHEP_mod}
\bibliography{w0_scale}

\providecommand{\href}[2]{#2}\begingroup\raggedright\begin{thebibliography}{10}

\bibitem{Sommer:1993ce}
R.~Sommer, {\it {A New way to set the energy scale in lattice gauge theories
  and its applications to the static force and alpha-s in SU(2) Yang-Mills
  theory}},  {\em Nucl.Phys.} {\bf B411} (1994) 839--854,
  [\href{http://xxx.lanl.gov/abs/hep-lat/9310022}{{\tt hep-lat/9310022}}].

\bibitem{Bernard:2000gd}
C.~W. Bernard, T.~Burch, K.~Orginos, D.~Toussaint, T.~A. DeGrand, et~al., {\it
  {The Static quark potential in three flavor QCD}},  {\em Phys.Rev.} {\bf D62}
  (2000) 034503, [\href{http://xxx.lanl.gov/abs/hep-lat/0002028}{{\tt
  hep-lat/0002028}}].

\bibitem{Donnellan:2010mx}
M.~Donnellan, F.~Knechtli, B.~Leder, and R.~Sommer, {\it {Determination of the
  Static Potential with Dynamical Fermions}},  {\em Nucl.Phys.} {\bf B849}
  (2011) 45--63, [\href{http://xxx.lanl.gov/abs/1012.3037}{{\tt
  arXiv:1012.3037}}].

\bibitem{Hasenfratz:2001hp}
A.~Hasenfratz and F.~Knechtli, {\it {Flavor symmetry and the static potential
  with hypercubic blocking}},  {\em Phys.Rev.} {\bf D64} (2001) 034504,
  [\href{http://xxx.lanl.gov/abs/hep-lat/0103029}{{\tt hep-lat/0103029}}].

\bibitem{Aoki:2010dy}
{\bf RBC Collaboration, UKQCD Collaboration}, Y.~Aoki et~al., {\it {Continuum
  Limit Physics from 2+1 Flavor Domain Wall QCD}},  {\em Phys.Rev.} {\bf D83}
  (2011) 074508, [\href{http://xxx.lanl.gov/abs/1011.0892}{{\tt
  arXiv:1011.0892}}].

\bibitem{Davies:2009tsa}
{\bf HPQCD Collaboration}, C.~Davies, E.~Follana, I.~Kendall, G.~Lepage, and
  C.~McNeile, {\it {Precise determination of the lattice spacing in full
  lattice QCD}},  {\em Phys.Rev.} {\bf D81} (2010) 034506,
  [\href{http://xxx.lanl.gov/abs/0910.1229}{{\tt arXiv:0910.1229}}].

\bibitem{Bazavov:2011aa}
{\bf Fermilab Lattice and MILC Collaborations}, A.~Bazavov et~al., {\it {B- and
  D-meson decay constants from three-flavor lattice QCD}},
  \href{http://xxx.lanl.gov/abs/1112.3051}{{\tt arXiv:1112.3051}}. 63 pages, 13
  figures.

\bibitem{Follana:2007uv}
{\bf HPQCD Collaboration, UKQCD Collaboration}, E.~Follana, C.~Davies,
  G.~Lepage, and J.~Shigemitsu, {\it {High Precision determination of the pi,
  K, D and D(s) decay constants from lattice QCD}},  {\em Phys.Rev.Lett.} {\bf
  100} (2008) 062002, [\href{http://xxx.lanl.gov/abs/0706.1726}{{\tt
  arXiv:0706.1726}}].

\bibitem{Kronfeld:2009cf}
A.~S. Kronfeld, {\it {The $f_{D_s}$ Puzzle}},
  \href{http://xxx.lanl.gov/abs/0912.0543}{{\tt arXiv:0912.0543}}.

\bibitem{Luscher:2009eq}
M.~Luscher, {\it {Trivializing maps, the Wilson flow and the HMC algorithm}},
  {\em Commun.Math.Phys.} {\bf 293} (2010) 899--919,
  [\href{http://xxx.lanl.gov/abs/0907.5491}{{\tt arXiv:0907.5491}}].

\bibitem{Narayanan:2006rf}
R.~Narayanan and H.~Neuberger, {\it {Infinite N phase transitions in continuum
  Wilson loop operators}},  {\em JHEP} {\bf 0603} (2006) 064,
  [\href{http://xxx.lanl.gov/abs/hep-th/0601210}{{\tt hep-th/0601210}}].

\bibitem{Luscher:2010iy}
M.~Luscher, {\it {Properties and uses of the Wilson flow in lattice QCD}},
  {\em JHEP} {\bf 1008} (2010) 071,
  [\href{http://xxx.lanl.gov/abs/1006.4518}{{\tt arXiv:1006.4518}}].

\bibitem{Luscher:2011bx}
M.~Luscher and P.~Weisz, {\it {Perturbative analysis of the gradient flow in
  non-abelian gauge theories}},  {\em JHEP} {\bf 1102} (2011) 051,
  [\href{http://xxx.lanl.gov/abs/1101.0963}{{\tt arXiv:1101.0963}}].

\bibitem{Durr:2010aw}
S.~Durr, Z.~Fodor, C.~Hoelbling, S.~Katz, S.~Krieg, et~al., {\it {Lattice QCD
  at the physical point: Simulation and analysis details}},  {\em JHEP} {\bf
  1108} (2011) 148, [\href{http://xxx.lanl.gov/abs/1011.2711}{{\tt
  arXiv:1011.2711}}].

\bibitem{Durr:2010vn}
S.~Durr, Z.~Fodor, C.~Hoelbling, S.~Katz, S.~Krieg, et~al., {\it {Lattice QCD
  at the physical point: light quark masses}},  {\em Phys.Lett.} {\bf B701}
  (2011) 265--268, [\href{http://xxx.lanl.gov/abs/1011.2403}{{\tt
  arXiv:1011.2403}}].

\bibitem{Durr:2011ap}
S.~Durr, Z.~Fodor, C.~Hoelbling, S.~Katz, S.~Krieg, et~al., {\it {Precision
  computation of the kaon bag parameter}},  {\em Phys.Lett.} {\bf B705} (2011)
  477--481, [\href{http://xxx.lanl.gov/abs/1106.3230}{{\tt arXiv:1106.3230}}].

\bibitem{Aoki:2006we}
Y.~Aoki, G.~Endrodi, Z.~Fodor, S.~Katz, and K.~Szabo, {\it {The Order of the
  quantum chromodynamics transition predicted by the standard model of particle
  physics}},  {\em Nature} {\bf 443} (2006) 675--678,
  [\href{http://xxx.lanl.gov/abs/hep-lat/0611014}{{\tt hep-lat/0611014}}].

\bibitem{Aoki:2006br}
Y.~Aoki, Z.~Fodor, S.~Katz, and K.~Szabo, {\it {The QCD transition temperature:
  Results with physical masses in the continuum limit}},  {\em Phys.Lett.} {\bf
  B643} (2006) 46--54, [\href{http://xxx.lanl.gov/abs/hep-lat/0609068}{{\tt
  hep-lat/0609068}}].

\bibitem{Aoki:2009sc}
Y.~Aoki, S.~Borsanyi, S.~Durr, Z.~Fodor, S.~D. Katz, et~al., {\it {The QCD
  transition temperature: results with physical masses in the continuum limit
  II.}},  {\em JHEP} {\bf 0906} (2009) 088,
  [\href{http://xxx.lanl.gov/abs/0903.4155}{{\tt arXiv:0903.4155}}].

\bibitem{Borsanyi:2010bp}
{\bf Wuppertal-Budapest Collaboration}, S.~Borsanyi et~al., {\it {Is there
  still any $T_c$ mystery in lattice QCD? Results with physical masses in the
  continuum limit III}},  {\em JHEP} {\bf 1009} (2010) 073,
  [\href{http://xxx.lanl.gov/abs/1005.3508}{{\tt arXiv:1005.3508}}].

\bibitem{Borsanyi:2010cj}
S.~Borsanyi, G.~Endrodi, Z.~Fodor, A.~Jakovac, S.~D. Katz, et~al., {\it {The
  QCD equation of state with dynamical quarks}},  {\em JHEP} {\bf 1011} (2010)
  077, [\href{http://xxx.lanl.gov/abs/1007.2580}{{\tt arXiv:1007.2580}}].

\bibitem{Endrodi:2011gv}
G.~Endrodi, Z.~Fodor, S.~Katz, and K.~Szabo, {\it {The QCD phase diagram at
  nonzero quark density}},  {\em JHEP} {\bf 1104} (2011) 001,
  [\href{http://xxx.lanl.gov/abs/1102.1356}{{\tt arXiv:1102.1356}}].

\bibitem{Borsanyi:2011sw}
S.~Borsanyi, Z.~Fodor, S.~D. Katz, S.~Krieg, C.~Ratti, et~al., {\it
  {Fluctuations of conserved charges at finite temperature from lattice QCD}},
  {\em JHEP} {\bf 1201} (2012) 138,
  [\href{http://xxx.lanl.gov/abs/1112.4416}{{\tt arXiv:1112.4416}}]. 13 pages,
  8 figures in Jhep style.

\bibitem{Durr:2008rw}
S.~Durr, Z.~Fodor, C.~Hoelbling, R.~Hoffmann, S.~Katz, et~al., {\it {Scaling
  study of dynamical smeared-link clover fermions}},  {\em Phys.Rev.} {\bf D79}
  (2009) 014501, [\href{http://xxx.lanl.gov/abs/0802.2706}{{\tt
  arXiv:0802.2706}}].

\bibitem{Durr:2008zz}
S.~Durr, Z.~Fodor, J.~Frison, C.~Hoelbling, R.~Hoffmann, et~al., {\it
  {Ab-Initio Determination of Light Hadron Masses}},  {\em Science} {\bf 322}
  (2008) 1224--1227, [\href{http://xxx.lanl.gov/abs/0906.3599}{{\tt
  arXiv:0906.3599}}].

\bibitem{Borsanyi:2011xx}
S.~Borsanyi et~al., {\it {Anisotropic lattices without tears}},  {\em
  WUB/12-03} (2012).

\bibitem{Edwards:2004sx}
{\bf SciDAC Collaboration, LHPC Collaboration, UKQCD Collaboration}, R.~G.
  Edwards and B.~Joo, {\it {The Chroma software system for lattice QCD}},  {\em
  Nucl.Phys.Proc.Suppl.} {\bf 140} (2005) 832,
  [\href{http://xxx.lanl.gov/abs/hep-lat/0409003}{{\tt hep-lat/0409003}}].

\bibitem{Shintani:2010ph}
E.~Shintani, S.~Aoki, H.~Fukaya, S.~Hashimoto, T.~Kaneko, et~al., {\it {Strong
  coupling constant from vacuum polarization functions in three-flavor lattice
  QCD with dynamical overlap fermions}},  {\em Phys.Rev.} {\bf D82} (2010)
  074505, [\href{http://xxx.lanl.gov/abs/1002.0371}{{\tt arXiv:1002.0371}}].

\bibitem{McNeile:2010ji}
C.~McNeile, C.~Davies, E.~Follana, K.~Hornbostel, and G.~Lepage, {\it
  {High-Precision c and b Masses, and QCD Coupling from Current-Current
  Correlators in Lattice and Continuum QCD}},  {\em Phys.Rev.} {\bf D82} (2010)
  034512, [\href{http://xxx.lanl.gov/abs/1004.4285}{{\tt arXiv:1004.4285}}].

\bibitem{Luscher:1984xn}
M.~Luscher and P.~Weisz, {\it {On-Shell Improved Lattice Gauge Theories}},
  {\em Commun.Math.Phys.} {\bf 97} (1985) 59.

\bibitem{Capitani:2006ni}
S.~Capitani, S.~Durr, and C.~Hoelbling, {\it {Rationale for UV-filtered clover
  fermions}},  {\em JHEP} {\bf 0611} (2006) 028,
  [\href{http://xxx.lanl.gov/abs/hep-lat/0607006}{{\tt hep-lat/0607006}}].

\bibitem{Durr:2010hr}
S.~Durr, Z.~Fodor, C.~Hoelbling, S.~Katz, S.~Krieg, et~al., {\it {The ratio
  FK/Fpi in QCD}},  {\em Phys.Rev.} {\bf D81} (2010) 054507,
  [\href{http://xxx.lanl.gov/abs/1001.4692}{{\tt arXiv:1001.4692}}].

\bibitem{Morningstar:2003gk}
C.~Morningstar and M.~J. Peardon, {\it {Analytic smearing of SU(3) link
  variables in lattice QCD}},  {\em Phys.Rev.} {\bf D69} (2004) 054501,
  [\href{http://xxx.lanl.gov/abs/hep-lat/0311018}{{\tt hep-lat/0311018}}].

\bibitem{Borsanyi:2011zza}
S.~Borsanyi, G.~Endrodi, Z.~Fodor, A.~Jakovac, S.~Katz, et~al., {\it {QCD
  equation of state from the lattice}},  {\em AIP Conf.Proc.} {\bf 1343} (2011)
  519--521.

\bibitem{Wolff:2003sm}
{\bf ALPHA collaboration}, U.~Wolff, {\it {Monte Carlo errors with less
  errors}},  {\em Comput.Phys.Commun.} {\bf 156} (2004) 143--153,
  [\href{http://xxx.lanl.gov/abs/hep-lat/0306017}{{\tt hep-lat/0306017}}].

\bibitem{Egri:2006zm}
G.~I. Egri, Z.~Fodor, C.~Hoelbling, S.~D. Katz, D.~Nogradi, et~al., {\it
  {Lattice QCD as a video game}},  {\em Comput.Phys.Commun.} {\bf 177} (2007)
  631--639, [\href{http://xxx.lanl.gov/abs/hep-lat/0611022}{{\tt
  hep-lat/0611022}}].

\end{thebibliography}\endgroup

\end{document}